\documentclass[aps,twocolumn,amsmath,amssymb,amsfonts]{revtex4}
\begin{document}
\title{Phase locking in quantum and classical oscillators: polariton condensates, lasers, and arrays of Josephson junctions}
\author{P. R. Eastham, M. H. Szymanska and P. B. Littlewood}
\affiliation{Cavendish Laboratory, Madingley Road, Cambridge, CB3 0HE. United Kingdom.}
\date{\today}

\begin{abstract}
We connect three phenomena in which a coherent electromagnetic field
could be generated: polariton condensation, phase-locking in arrays of
underdamped Josephson junctions, and lasing. All these phenomena have
been described using Dicke-type models of spins coupled to a single
photon mode. These descriptions may be distinguished by whether the
spins are quantum or classical, and whether they are strongly or
weakly damped. 
\end{abstract}


\maketitle

\section{Introduction}

Phase-locking\cite{pikovsky2001} of coupled oscillators is a
well-known phenomenon in nonlinear dynamics. The generation of
coherent radiation from Josephson junction arrays\cite{jain1984} is
but one example for macroscopic oscillators. But phase-locking exists
not only in classical systems but in quantum models.  Here there are
two basic paradigms for coherence of microscopic oscillators:
Bose-Einstein condensation (BEC), which is responsible for
superfluidity and superconductivity, and lasing.

Although BEC, lasing, and classical phase-locking all involve
collective coherent behaviour, they are usually described in very
different terms. With some exceptions\cite{borenstein1972},
descriptions of BEC and lasing are given in quantum-mechanical
language, which gives the impression that these phenomena derive from
quantum mechanics. Descriptions of Bose condensation sometimes go
further, suggesting that the condensate itself is a quantum mechanical
object. Nevertheless, it is unclear how these phenomena differ from
classical phase-locking.

The aim of this paper is to clarify the relationships amongst BEC,
lasing, and classical phase-locking, and hence the extent to which one
can describe the coherence in Bose condensates and lasers as
``quantum'' or ``classical''. To do this, we will consider Bose
condensation of cavity polaritons, phase-locking in arrays of coupled
Josephson junctions, and lasing. The simplest theoretical models of
these phenomena are in fact similar, and admit some controlled
solutions, enabling us to compare them cleanly.

\section{Cavity Polariton Condensation}

A cavity polariton\cite{cavpol,expolreview2} is the quantum of the
electromagnetic field in an optical cavity containing a dielectric. It
is the confined version of the bulk polariton considered many years
ago by Hopfield\cite{hopfield1958}, which is formed from propagating
photons coupled to electronic excitations such as excitons. Since
polaritons are photons coupled to other excitations they are bosons,
and therefore might be candidates for Bose condensation.

While the idea of a Bose condensate of bulk polaritons has been
discussed for many years\cite{keldyshbec1995,ex-bec-book}, it would be
an unusual type of condensate. This is because low-energy polaritons
are merely long-wavelength photons, which are not conserved
particles. Thus the polaritons cannot condense in the ground state,
making the bulk polariton condensate an intrinsically non-equilibrium
phenomenon. However, a condensate of cavity polaritons is not
necessarily a non-equilibrium phenomenon, because the lifetime of the
low-energy cavity polaritons is finite. If this lifetime were long
compared with the thermalisation time, one could consider the thermal
equilibrium of a population of polaritons, treated as conserved
particles. This is the normal situation for Bose condensation.

Eastham and Littlewood\cite{sscpaper,thesis,prbpaper} have considered
such quasi-equilibrium polariton condensation in a toy model. The
model is appropriate to localised electronic excitations, such as
excitons bound to impurities or localised on quantum dots, dipole
coupled to a single mode of a three-dimensional cavity. It has the
Hamiltonian
\begin{equation}
\label{eq:spinham} H = \omega_c \psi^\dagger \psi + \sum_i \left[E_i S^{z}_{i} +
\frac{g}{\sqrt{N}} \left( S^{+}_{i} \psi + \psi^\dagger S^-_i \right)
\right]. \end{equation} $\psi^\dagger$ is the creation operator for a
cavity photon, with energy $\omega_c$. The dielectric is modelled as a
set of $N$ two-level systems, with the $i^{th}$ two-level system
described by the spin-half operators $\vec{S_i}$. The eigenstates of
$S_i^z$ correspond to the presence or absence of an excitation on site
$i$.

The model (\ref{eq:spinham}) is the Dicke model\cite{dickemodel} of
quantum optics, which has been studied in many different regimes. To
apply it to polariton condensation one should find its ground state
or partition function, fixing the total number of excitations
\begin{displaymath} N_{ex}=\psi^\dagger\psi+\sum_i
\left(S_i^z+\frac{1}{2}\right), \end{displaymath} which is conserved
by the Hamiltonian (\ref{eq:spinham}). The constraint on
$N_{\mathrm{ex}}$ can be dealt with by introducing a chemical
potential $\mu$, so that one works with the unconstrained effective
Hamiltonian $H_{\mathrm{eff}}=H-\mu N_{ex}$. $H_{\mathrm{eff}}$ is the
same as $H$, except that the photon and exciton energies $\omega_c$
and $E_i$ are shifted by the chemical potential.

One can write down the ground-state wavefunction of $H_{\mathrm{eff}}$
by generalising the standard wavefunction for a Bose condensate. For
bosons with creation operator $b^\dagger$, the ground-state is the
coherent state \begin{equation} \label{eq:cohstate} e^{\lambda
b^\dagger}|\mathrm{vac}\rangle. \end{equation} In general, polaritons
are superpositions of an excitation of the cavity mode and an
excitation of the dielectric. Thus the generalisation of
(\ref{eq:cohstate}) to describe polariton condensation is
\begin{equation} \label{eq:polcohstate} \exp(\lambda\psi^\dagger + \sum_i e^{i\phi_i}w_i S_i^+)|\mathrm{vac}\rangle, \end{equation}
where
$\lambda$, $w_i$ and $\phi_i$ are variational parameters. Minimizing
$\langle H_{\mathrm{eff}}\rangle$ over these parameters gives an
equation for $\lambda$ which is analogous to the BCS gap equation:
\begin{equation}\label{eq:gap}(\omega_c-\mu)\lambda=\frac{g^2\lambda}{N}\sum_i
\frac{1}{\sqrt{(E_i-\mu)^2+4g^2|\lambda|^2}}.\end{equation}

The polariton condensate (\ref{eq:polcohstate}) is a superposition of
coherent states of the dielectric and the electromagnetic field. It
has a finite expectation value for the amplitude of the cavity field,
$\langle\psi\rangle$, and the electronic polarisation $\langle
S_i^-\rangle$.  The $\phi_i$ are the phase differences between the
electronic polarisation and the cavity field. They are fixed by the
dipole interaction term in (\ref{eq:spinham}), which is responsible
for the phase locking: it ensures that all the oscillators with a
finite polarisation are mutually coherent, $\phi_i=\phi$, when the
energy is minimised.

To make the connection to phase-locking more explicit, we note that
the gap equation is a special case of the condition for the dynamics
of the spins to synchronize at frequency $\mu$. In a frame rotating at
this frequency, the Heisenberg equations of motion corresponding to
the Hamiltonian (\ref{eq:spinham}) are
\begin{eqnarray} i\dot{\psi}&=&(\omega_c-\mu) \psi +
\frac{g}{\sqrt{N}}\sum_i S_i^- \label{eq:psimotion}\\ i\dot{S}_i^-&=&
(E_i-\mu) S_i^- - \frac{2g}{\sqrt{N}} S^{z}_i \psi \label{eq:s-motion} \\
i\dot{S}_i^z&=&\frac{g}{\sqrt{N}}\left(S_i^+\psi-\psi^\dagger
S^{-}_{i}\right). \label{eq:szmotion} \end{eqnarray} In an unsynchronized
state, the sum on the right-hand side of Eq.\ \ref{eq:psimotion} is of
order $\sqrt{N}$, so $\psi$ is of order $1$. For such a $\psi$ the
spins are free to leading order in $N$ due to the scaling of the
coupling constant. The spin on site $i$ simply precesses around the z
axis at its natural frequency $(E_i-\mu)$. In a synchronized state,
part of the sum in Eq.\ \ref{eq:psimotion} will be of order $N$, so
$\psi$ will be of order $\sqrt{N}$. Such a field gives a finite
contribution to the effective magnetic field on each spin. The
dynamics of $\langle S_i^- \rangle$ in this effective field contains a
static component. Substituting this static component into
(\ref{eq:psimotion}) and setting $\dot{\psi}=0$ gives a
self-consistency condition on the synchronized states,
\begin{equation}
\label{eq:gapsync}
(\omega_c-\mu)\lambda=\frac{g^2\lambda}{N}\sum_i
\frac{2\langle S^{z\prime}_i\rangle_0}{\sqrt{(E_i-\mu)^2+4g^2|\lambda|^2}}.
\end{equation}
This condition is a generalisation of (\ref{eq:gap}), in which the
unit numerator in the $i^{th}$ term of the sum becomes $2\langle
S^{z\prime}_i \rangle_0$. $S^{z\prime}_i$ is the component of spin $i$
along its effective field, and $\langle \rangle_0$ denotes an
expectation value in the initial state. The particular choice of
``initial conditions'' that appears in Eq.\ \ref{eq:gap} corresponds
to a thermal (here T=0) occupation of the quasiparticle states,
producing a solution of Eq.\ \ref{eq:spinham} of the lowest free energy.

The self-consistent approach suggests that a mean-field theory in the
amplitude of the cavity mode is exact as $N\to\infty$. This can be
formally demonstrated by constructing the partition function as a path
integral, which can be evaluated using saddle-point
techniques. Physically, the mean-field theory is exact because the
cavity mode is coupled to many electronic states, and so should have
relatively small fluctuations. A consequence is that condensation in
the model (\ref{eq:spinham}) is, in the limit of a large system, no
more than the phase-locking of {\em classical} coupled oscillators.


\section{Josephson Junction Arrays}
\label{sec:jj}

A different system which can be described by models similar to
(\ref{eq:spinham}) is a Josephson junction array in a microwave
cavity. Phase-locking in that system was considered in 1970 by
Tilley\cite{tilley1970}, in a model describing junctions connected in
series in a single-mode cavity. He considered the fully synchronized
states of the array, in which each junction oscillates at the same
frequency. While there are many such states, differing in locking
frequency, phase configuration and photon number, he showed that one
particular state is selected by the driving current and cavity
losses.

To compare locking in Josephson junctions to polariton condensation,
we use the angular momentum representation\cite{leggett2001} of a
Josephson junction. Each junction is represented by a spin $S$, whose
magnitude is half the total number of condensed Cooper pairs in the
two superconductors connected by the junction. $S^z$ is half the
difference between the number of condensed pairs on either side of the
junction, while $S^+$ and $S^-$ transfer condensed pairs across the
junction. These operators obey the usual angular momentum commutation
rules to the extent that Cooper pairs can be treated as structureless
bosons, which should be a good approximation in the weak-coupling
limit $\Delta\ll\omega_D$. Note that $S$ is usually much larger than
$1$ --- the junctions are themselves macroscopic objects.

\begin{widetext}
In the angular momentum representation, the canonical Hamiltonian for
$N$ Josephson junctions interacting with a microwave resonance
is\cite{bonificio1982,lee1971}\footnote{Note that the normalisation
used for the spin operators in Refs. \onlinecite{bonificio1982} and
\onlinecite{lee1971} differs from that required by our
definitions. For example, $S^z$ in Ref. \onlinecite{lee1971} is the
difference in density, rather than number, of pairs.}
\begin{equation}\label{eq:jjspinham} H = \omega_c \psi^\dagger \psi +
\sum_i \left[K_i (S^{z}_{i })^2 + \frac{g_i}{\sqrt{N}} \left( S^{+}_{i} - S^-_i
\right)\left(\psi-\psi^\dagger\right) + j_i (S_i^++S_i^-) + \nu_i
S^z_i \right]. \end{equation} $j_i$ corresponds to the standard
Josephson tunnelling, $g_i$ to the photon assisted tunnelling, $K_i$
to the charging energy of the junction, and $\nu_i$ to a voltage bias
across the junction. To bring (\ref{eq:jjspinham}) as close as
possible to the polariton condensation problem, we neglect the
standard Josephson couplings $j_i$ and make the rotating-wave
approximation. This gives the Hamiltonian
\begin{equation}\label{eq:jjspinhamrwa} H_{\mathrm{JJ}}\approx \omega_c \psi^\dagger \psi +
\sum_i \left[K_i (S^{z}_{i })^2 + \frac{g_i}{\sqrt{N}} \left(
S^{+}_{i} \psi + \psi^{\dagger} S^-_i \right) + \nu_i S^z_i
\right]. \end{equation} Since (\ref{eq:jjspinhamrwa}) conserves
$N_{\mathrm{ex}}$, we may consider its behaviour at fixed
$N_{\mathrm{ex}}$\cite{bonificio1982}. This problem should be exactly
solvable in the limit $N\to\infty$ using a mean-field theory in the
amplitude of the cavity mode.

To construct the mean-field theory for
$H^\prime_{\mathrm{JJ}}=H_{\mathrm{JJ}}-\mu N_{\mathrm{ex}}$, we
should follow the same self-consistent approach that was used for
polariton condensation. This is complicated, however, by the presence
of the charging energy. Because of this term, the single-spin
effective Hamiltonian of the mean-field theory is nonlinear. Instead
of directly tackling this problem, we will follow the variational
procedure used for polariton condensation. While this corresponds
exactly to the mean-field theory for the polariton condensate, it will
only be an approximation to the mean-field theory of the array. This
is because (\ref{eq:polcohstate}) is the ground state of the effective
Hamiltonian which occurs in the mean-field theory of polariton
condensation, but not of that which occurs for the array.

Looking for stationary points of $\langle H^\prime_{\mathrm{JJ}}
\rangle$ in the variational state (\ref{eq:polcohstate}) gives
equations for $\lambda$ and $\theta_i=\arctan w_i$:
\begin{eqnarray}
(\omega-\mu)|\lambda|&=&\sum_i g^{\prime}_{i} \sin(2\theta_i),
\label{eq:jjgap1} \\ \sin(2\theta_i)((\mu^\prime-\nu_i^\prime)+2
K_i^\prime S \cos(2\theta_i)) &=& - 2 g^{\prime}_{i}
|\lambda|\cos(2\theta_i). \label{eq:jjgap2}
\end{eqnarray} Here primes denote scaled variables, $\mu^\prime=\mu
S$ etc., and the minimum energy solution has $\sin(2\theta_i)>0$. For
some $N_{\mathrm{ex}}$ and $K_i$, the charging energy term in
(\ref{eq:jjgap2}) will be negligible. Eqs.\ \ref{eq:jjgap1} and \
\ref{eq:jjgap2} are then just the gap equation (\ref{eq:gap}) derived
for polariton condensation, with the replacements $\frac{1}{2}g\to g^\prime$ and
$\frac{1}{2}(E_i-\mu)\to (\nu^\prime-\mu^\prime)$.

\end{widetext}

We have not investigated the consequences of the factors of $S$ and
the charging energy in
(\ref{eq:jjgap1}--\ref{eq:jjgap2}). Nevertheless, it seems that the
mean-field theories for phase-locking in a Josephson array and for
polariton condensation are very similar.  This is perhaps surprising,
because we tend to think of Bose condensation as a truly quantum
phenomenon occurring for microscopic oscillators, while Josephson
junctions are macroscopic ($S \gg 1$), so that phase-locking is
naturally thought of in terms of classical coupled
oscillators. However as we stressed in the last section, at zero
temperature quantum mechanics is irrelevant to the {\em mean-field}
theory of polariton condensation: the form of the gap equation
(\ref{eq:gap}) is the same for quantum spins and for classical angular
momenta. The reason for this can be seen in the self-consistent
dynamical approach, in which the problem is reduced to that of spins
in a self-consistent field. For such a linear problem, the
commutativity or otherwise of the spin components is
irrelevant. Notice however that there are at least two routes to the
classical limit: in the Josephson array, the individual elements
become classical as $S \to \infty$, whereas for the polariton
condensate we have $S=1/2$, but $N \to \infty$. In the latter system,
only the coherent ground state can be treated as a classical object.


\section{Decoherence of the polariton condensate}

Some of the conspicuous differences between quantum and classical
oscillators are due to the decoherence of quantum oscillators by their
environment. Unlike a classical oscillator, a quantum oscillator has
states which do not have a well-defined phase. Furthermore, we expect
that the environment will drive it towards such states. Thus we might
expect that interactions with the environment would have a significant
effect on the phase-locking of quantum oscillators. In fact, we shall
see that an infinite condensate is immune to weak decoherence
phenomena, in the same way that a superconductor is immune to weak
phase-breaking. But in the case of strong decoherence, and perhaps in
the case of a finite system, we will find a connection to yet another
example of macroscopic phase-locking --- the laser.


The laser and the polariton condensate are usually studied in separate
contexts, and the connection between them is not made. This should be
surprising, as both can be described by {\em exactly the same}
Hamiltonian \eqref{eq:spinham}. However, in a conventional
laser\cite{hakenbook} the only significant ordering is the coherence
of the photons, while in the polariton condensate both the photons and
the excitons are coherent. This is because in a conventional laser the
polarisation of the gain medium, $\langle S^+\rangle$, is strongly
damped by processes such as pumping, collisions, and interactions with
phonons and impurities. The coherence in the photons remains, because
it can be generated by stimulated emmission even from an incoherent
reservoir. 



The effects on the polariton condensate of different kinds of
decoherence processes have recently been studied by two of us
\cite{mhs2002,mhs,mhs2003}, using models related to \eqref{eq:spinham}. These
models are obtained by rewriting each spin operator in terms of a pair
of fermions, with annihilation operators $a_i$ and $b_i$. This is done
by replacing $S_i^+$ with $b^\dagger_ia_i$ and $S_i^z$ with
$\frac{1}{2}(b^\dagger_i b_i - a^\dagger_i a_i)$. With the local
constraints $b^\dagger_i b_i + a^\dagger_i a_i=1$ this would give an
exact representation of the model \eqref{eq:spinham}. In our studies
of decoherence, however, we replace these local constraints with their
global equivalent. The decoherence is modelled using baths of harmonic
oscillators. Thus we consider the Hamiltonian
\begin{equation}
H = H_S+H_{SB}+H_{B}.
\label{Hstart}
\end{equation}
The first term $H_S$ is just the Hamiltonian \eqref{eq:spinham}
written in terms of the fermionic operators,
\begin{equation}
H_S = \omega_c
\psi^\dagger\psi + \sum_{i=1}^N \frac{E_i}{2}
(b^\dagger_ib_i-a^\dagger_ia_i) +
\frac{g_i}{\sqrt{N}}(b^\dagger_ia_i\psi + \psi^\dagger
a^\dagger_ib_i).
\label{HS}
\end{equation}
$H_{B}$ is a quadratic Hamiltonian describing the baths, and $H_{SB}$
describes the coupling between the system and the baths. The most
general form of $H_{SB}$ is
\begin{eqnarray}
H_{SB}&=&\sum_{k} g^{\kappa}_{k}(\psi^\dagger d_{k}  +
d_{k}^\dagger \psi) \nonumber \\
&+&
\sum_{i,k}[g^{\gamma_{\uparrow}}_{i,k}
(b^\dagger_ia_i c_{k}^{\alpha\dagger}+c_{k}^{\alpha}
a^\dagger_ib_i) \nonumber \\ &+&
g^{\gamma_{\downarrow}}_{i,k}
(b^\dagger_ia_i c_{k}^{\beta}+
c_{k}^{\beta \dagger}
a^\dagger_ib_i)  \nonumber \\
&+&\Gamma^{(1)}_{k}(b^\dagger_ib_i
+a^\dagger_ia_i)(c_{k}^{\zeta\dagger} +c_{k}^\zeta) \nonumber \\ &+&
\Gamma^{(2)}_{i,k}(b^\dagger_ib_i
-a^\dagger_ia_i)(c_{k}^{\theta\dagger} +c_{k}^\theta)].
\label{eq:HSB}
\end{eqnarray}

The first term in (\ref{eq:HSB}) describes the decay of the cavity
mode, the second term pumping of the two-level oscillators, while the
third term contains all the processes which destroy the electronic
excitations, such as sponteneous emmission into modes other than the
cavity mode. These baths could give rise to a flow of excitation
through the system. The fourth and the fifth terms, however, describe
all the dephasing processes which do not change the total number of
excitations in the cavity, for example collisions and interactions
with phonons and impurities. Processes described by the second, the
third and the fourth terms in (\ref{eq:HSB}) have pair-breaking
character, analogous to magnetic impurities in superconductors, and
correspond to potentials which vary rapidly in space or in time.

%
%

In order to establish a crossover between an isolated condensate and a
laser the decoherence processes must be included
self-consistently. The widely used quantum Maxwell-Bloch (Langevin)
equations with a constant decay rate for the polarisation are not
correct when the coherent polarisation is large, i.e. for the
polariton condensate. In these equations the collective behaviour of
the excitons is not taken into account when the lifetime for
polarisation is derived. Instead, the lifetime for a single exciton is
used in the equation for a collective polarisation mode. 

To treat decoherence processes self-consistently, we use a procedure
analogous to the Abrikosov-Gor'kov theory\cite{abrikosov1961} of
magnetic superconductors. In this theory, the baths which model
decoherence are integrated out, introducing effective interactions
between different two-level systems. These interactions are expressed
as a self-energy in Dyson's equation, $G^{-1}_{ij} =
G_{0,ij}^{-1}-\Sigma_{ij}$, of the form $\Sigma_{ij} = \gamma G_{ij}$. This
form should be contrasted with the non self-consistent treatment, in
which the decoherence appears as a constant lifetime in the Dyson
equation.


It turns out that the phase-locked polariton condensate is a robust
phenomenon because at low decoherence strength it is protected by an
energy gap proportional to the photon field amplitude. This gap
becomes smaller and finally gets suppressed as the pair-breaking
decoherence is increased. At low excitation densities this leads to
the suppression of all the coherent fields while at high densities it
leads instead to the conventional characteristic of a semiconductor
``laser'' in which the coherence is almost entirely in the photon
field and there is no gap in the excitation spectrum. The laser regime
of a polariton system emerges in a way that demonstrates its close
analogy to a gapless superconductor.

Although the coherent polarisation in a conventional laser is strongly
damped, it must be finite for the laser to operate. Thus the
transition between a condensate and a laser is smooth. There is no
formal distinction between the two based on the broken symmetry of the
ground state. There may be useful practical distinctions however, such
as the presence or absence of a gap in the excitation spectrum. We do
not yet know if the dynamics of the order parameter, and hence the
linewidth, differs. Note that in the Abrikosov-Gor'kov theory the
pair-breaking does not produce fluctuations in the order
parameter. The theory presented here may share this feature, whereas
real lasers have a finite linewidth.


\section{Finite-size fluctuations}

The question of classical or quantum behaviour can never arise for the
dynamics of the order parameter in an (infinite) system with a broken
symmetry.  One simply has a macroscopic equation of motion for the
order parameter interacting with an external classical field --- with
a familiar example being the Josephson equation for a weak link, and a
less familiar one the classical dynamics of the (averaged) mean field
equations in section II.  However, when such a system becomes finite
(though still large) in extent, we can ask whether the dynamics are
now best descibed by a Schrodinger equation or a Langevin equation:
the order parameter will ``diffuse'', at short times following quantum
mechanics, and at longer times dictated by Brownian motion.  Of
course, we are now concerned only with the low energy degrees of
freedom --- those near the frequency $\mu$ in the rotating frame ---
and certainly well away from the quasiparticle excitations above the
gap.

The procedure to be followed is clear, at least in principle, though
it has not been fully completed for the model of a polariton
condensate.  As we mentioned briefly above, the variational equations
correspond to the saddle point of a quantum mechanical action, which
is exact as $N\to\infty$. Fluctuations at finite $N$ are described by
a new effective action, with degrees of freedom that are then coupled
to baths exactly as in Eq.\ref{eq:HSB}.  However, in contrast to the
results of the last section, phase-breaking perturbations are expected
to be {\em always relevant} however weak.  We note in passing that
this methodology is different from the conventional procedure to begin
with a classical action that is then re-quantised. Whether or not it
yields any distinct difference is not known.

For a finite system, the broken symmetry ground state will not be
stable, and we will observe fluctuations; this is a familiar point of
view in classical laser theory, where Haken has emphasised how the
mean field theory corresponds to a second-order phase transition, and
the fluctuations in a real system arise because the number of photons
is not infinite. However, with the starting point of an effective
quantum mechanical action one no longer presupposes a classical limit.
With the new action one can compute, for example, correlation
functions of the photon field, which are directly measurable. At the
mean field level, the photon field is a classical electromagnetic
field, so the distinction between quantum and classical statistics
will only appear at this stage.
 
\section{Discussion}

The distinctions between coupled oscillators, BEC of polaritons, and
strongly damped lasers is an important one in the context of recent
experiments on semiconductor microcavities. We will not give a
detailed review of the field here, except to point out that one recent
experiment \cite{deng2002} has shown evidence for coherence of the
photon field in the nonlinear but incoherently pumped
microcavity. However, if we define BEC of polaritons to be restricted
to systems where the excitonic degrees of freedom have strong
coherence, observation of photon coherence is by itself not decisive
evidence of polariton BEC.  One characteristic of this regime would be
a gap in the excitation spectrum, which was not apparently observed.


The difficulties of making these distinctions had already arisen in
the Josephson junction array problem --- here embedded in microwave
cavities. As we discussed in section\ \ref{sec:jj}, the Hamiltonian is
mathematically similar to that of the polariton
condensate. Recently\cite{barbara1999}, Barbara et al. reported
thresholds in the ac output power of such an array as the driving
power was increased. They interpreted their results as analogous to
lasing, with gain due to stimulated Josephson
tunnelling. Subsequently, Stroud et al. showed that many of the
experimental observations could be reproduced by classical
treatments\cite{stroudgroundstate2000,stroud2ddynamics2003,stroud1ddynamics2001,stroud1ddynamics2002}
of models similar to \eqref{eq:jjspinham}.

The phenomena of lasing and BEC are well-known to be closely
connected. At a microscopic level, both are consequences of the
quantum mechanics of indistinguishable bosons, and more specifically
of stimulated scattering. Lasing is described dynamically, and the
role of stimulated scattering is explicit. Bose condensation is
described thermodynamically, with the role of stimulated scattering
hidden in the Bose-Einstein distribution. On a more sophisticated
level, Haken\cite{haken1975} has shown that the mean-field theory of
the laser is analogous to that of a second-order phase transition,
while Oraevskii\cite{oraevskii1985} has discussed the dynamics of a
superconductor in terms of stimulated scattering.

What confuses the issue about BEC is that the conventional textbook
picture presents BEC as a consequence of statistical physics of weakly
or non- interacting bosons, which obscures the central point that BEC
in a macroscopic system is a phase transition like any other. So in a
very large system one will not expect to find quantum mechanics
operating at the level of the macroscopic order parameter, even if the
microscopic theory of this phase transition requires quantum physics.
In the JJ array, it is usual to imagine that the individual elements
are macroscopic (``decohered'') from their environment, but one can
see from the above that is is not necessary to assume this in order to
develop the correct classical theory of the phase locked coupled
array. For our simple model of polariton BEC one can see again that in
the large system limit with macroscopic occupancy, the quantum
mechanical ground state corresponds to the dynamics of classical
phase-locked oscillators.  But here there is a possibility to decohere
the individual elements (spin half dipoles) from each other (by
coupling to external baths) and restore conventional laser theory with
a coherent photon field, supported by incoherent electronic
polarisation. But still, in the limit of macroscopic occupation, the
coherent photon field is essentially a classical one.




Most interesting would be the behaviour of large, but not infinite,
systems. Here the order parameter fluctuates generically due to finite
size effects. These fluctuations can arise due to environmental
interactions, which will give rise to classical diffusion, as in the
standard theory of the laser linewidth near threshold. They might also
arise because the order parameter tunnels between equivalent states,
as has been achieved in systems of small Josephson
junctions\cite{mooij2000,friedman2000}. The competition between
quantum mechanical tunneling and environmental dephasing is of course
at the heart of current attempts to create quantum coherent devices
--- and if excitonic or polaritonic BEC were observed, this would
provide another possible fundamental system upon which to base such
work.


\section{Acknowledgements}

PRE and MHS acknowledge the support of research fellowships from
Sidney Sussex and Gonville and Caius colleges, Cambridge,
respectively.  This work is also supported by the EPSRC and by the EU
network ``Photon-mediated phenomena in semiconductor nanostructures''.



\end{document}